**A Quantum Mechanical Insight into S$_N$2 Reactions:**

**Semiclassical Initial Value Representation Calculations of Vibrational Features of**

**the Cl⁻---CH$_3$Cl Pre-Reaction Complex with the VENUS Suite of Codes**

Xinyou Ma[1], Giovanni Di Liberto[2], Riccardo Conte[2],

William L. Hase[1]*, Michele Ceotto[2]*

[1]Department of Chemistry and Biochemistry

Texas Tech University

Lubbock, Texas 79409, USA

[2]Dipartimento di Chimica

Università degli Studi di Milano

via C. Golgi 19, 20133 Milano, Italy

Email: bill.hase@ttu.edu; michele.ceotto@unimi.it

**Abstract**

The role of vibrational excitation of reactants in driving reactions involving polyatomic species has been often studied by means of classical or quasi-classical trajectory simulations. We propose a different approach based on investigation of vibrational features of the $Cl^-$---$CH_3Cl$ pre-reaction complex for the $Cl^-$ + $CH_3Cl$ $S_N2$ reaction. We present vibrational power spectra and frequency estimates for the title pre-reaction complex calculated at the level of classical, semiclassical, and second-order vibrational perturbation theory on a pre-existing analytical potential energy surface. The main goals of the paper are the study of anharmonic effects and understanding of vibrational couplings that permit energy transfer between the collisional kinetic energy and the internal vibrations of the reactants. We provide both classical and quantum pictures of intermode couplings, and show that the $S_N2$ mechanism is favored by coupling of a C-Cl bend involving the $Cl^-$ projectile with the $CH_3$ rocking motion of the target molecule. We also illustrate how the routines needed for semiclassical vibrational spectroscopy simulations can be interfaced in a user-friendly way to pre-existing molecular dynamics software. In particular, we present implementation of semiclassical spectroscopy into the VENUS suite of codes, thus providing a useful computational tool for users who are not experts of semiclassical dynamics.

**I. Introduction**

New quantum dynamics methods are important for the chemistry community to achieve an increasing accuracy in predicting and/or explaining experimental findings. However, often the use of quantum dynamics methods is confined within the laboratory that has introduced them. Instead, it is beneficial for any new method to be implemented in a way accessible to the greater community in order to test its robustness and provide useful feedback for further development of the approach.

In this paper we introduce a user-friendly implementation of the semiclassical initial value representation (SC-IVR) method[1-4] for spectroscopic calculations via the VENUS classical chemical dynamics suite of codes[5, 6]. We then test the new implementation by calculating the vibrational power spectrum of the nucleophilic pre-reaction complex $Cl^-$---$CH_3Cl$ of the $Cl^-$ + $CH_3Cl$ $S_N2$ reaction. In fact, a very important problem in gas-phase reaction dynamics is to understand how vibrational excitations of reactants affect the outcome of the chemical reaction. This problem is common to both bimolecular and unimolecular reactions and quite often it is not clear how mode selective vibrational excitations may enhance the reactivity of a pre-reaction complex.

The complex is shown in Figure 1. The bond between $Cl_a$ and the methyl group is much weaker than the $C$-$Cl_b$ bond, since $Cl_a$ is the anion colliding with $CH_3Cl_b$, forming the $Cl_a^-$---$CH_3Cl_b$ ion-dipole complex.

The goal of the paper is twofold. First, we want to explain how to better introduce the SC-IVR molecular dynamics (MD) into an existing molecular dynamics code running classical trajectories both on pre-fitted potential subroutines (VENUS[5, 6]) or using on-the-fly calculations (NWChem-VENUS[7, 8]). This approach can be applied to other existing chemical dynamics codes and semiclassical methods have shown a good balance between feasibility and quantum accuracy.[2,

[9-24] Secondly, we aim at describing how SC-IVR may be employed to shed some light on the mechanism of nucleophilic reactions from a quantum mechanical point of view. More specifically, we look at the importance of vibrational couplings in promoting an $S_N2$ reaction, and the role played by anharmonic effects, which are neglected by the commonly employed harmonic approximation.

Classical dynamics have been an important way to computationally study $S_N2$ reactions.[25-32] A major finding from the classical simulations is that only a few modes of the pre-reaction complex are active in intramolecular vibrational energy redistribution.[26] As a consequence, the kinetic energy of the $Cl_a^-$ projectile is redistributed preferentially among these few modes. This inefficient intra-vibrational energy redistribution, which is not statistical and does not follow Rice–Ramsperger–Kassel–Marcus (RRKM) theory, is an important component of the $S_N2$ reaction mechanism.[25, 30] A shortcoming of these simulations is that they provide a classical picture of the reaction dynamics without taking into account any quantum dynamical effects. Such dynamics have been partially accounted for by reduced dimensional quantum dynamics simulations.[33-46] In addition, a series of quasi-classical trajectories[47-53] have been performed to supplement the previous simulations. We move a step forward in including quantum mechanical effects in the study of $S_N2$ reactions by focusing on the $Cl^-$---$CH_3Cl$ pre-reaction complex. Specifically, we study the full-dimensional quantum vibrational motion by Fourier transforming the vibrational wavepacket autocorrelation function with the goal to decipher vibrational couplings between modes.

The paper is organized as follows. In Section II the SC-IVR method for spectroscopic calculations is recalled. In Section III we show how the method has been implemented into the VENUS suite of codes. Section IV illustrates some approaches which can be employed for studying the vibrational couplings, including a tailored one based on the SC-IVR method. The results are presented and discussed in Section V, while Section VI concludes the paper.

## II. The Semiclassical Initial Value Representation Method for Spectroscopic Calculations

In this paper we focus on the calculation of power spectra $I(E)$ of molecular systems. Given the Hamiltonian $\hat{H}$, its power spectrum is the Fourier transform of the survival probability amplitude[54] of an arbitrarily given reference state $|\chi\rangle$

$$I(E) \equiv \frac{1}{2\pi\hbar} \int_{-\infty}^{+\infty} \langle \chi | e^{-i\hat{H}t/\hbar} | \chi \rangle e^{iEt/\hbar} dt \qquad (1)$$

In the semiclassical (SC) approximation, the quantum time-evolution operator $e^{-i\hat{H}t/\hbar}$ appearing in eq. (1) is reformulated in a more practical way by stationary phase approximating the sum over all possible paths appearing in the Feynman Path Integral representation.[55] This restricts the sum over all possible paths to a sum over all possible classical trajectories connecting the two endpoints of the propagator.[56] Unfortunately, the rigid double-boundary condition makes the search for these classical trajectories cumbersome. W. H. Miller[1, 2, 57-59] greatly simplified this search by demonstrating that it could be substituted by a Monte Carlo initial phase space integration in an approach known as the SC Initial Value Representation method. Later, Heller first,[60-62] Herman and Kluk,[63, 64] and Kay later[65-67] reformulated the SC-IVR propagator adopting a coherent state representation,[54, 60, 61, 68] commonly called the Herman-Kluk propagator, in which the survival probability in eq. (1) becomes

$$\langle\chi|e^{-i\hat{H}t/\hbar}|\chi\rangle \approx \frac{1}{(2\pi\hbar)^F}\iint d\mathbf{p}(0)d\mathbf{q}(0)C_t(\mathbf{p}(0),\mathbf{q}(0))e^{\frac{i}{\hbar}S_t(\mathbf{p}(0),\mathbf{q}(0))}$$

$$\times \langle\chi|\mathbf{p}(t),\mathbf{q}(t)\rangle\langle\mathbf{p}(0),\mathbf{q}(0)|\chi\rangle \quad (2)$$

and where $(\mathbf{p}(0), \mathbf{q}(0))$ is the $F$-dimensional classical trajectory initial phase space condition, $S_t(\mathbf{p}(0), \mathbf{q}(0))$ is the classical action, and $C_t(\mathbf{p}(0), \mathbf{q}(0))$ indicates the pre-exponential factor. The coherent states are of the type

$$\langle\mathbf{x}|\mathbf{p}(t),\mathbf{q}(t)\rangle = \left(\frac{\det(\Gamma)}{\pi^F}\right)^{\frac{1}{4}} e^{-(\mathbf{x}-\mathbf{q}(t))^T \Gamma(\mathbf{x}-\mathbf{q}(t))/2 + i\mathbf{p}^T(t)(\mathbf{x}-\mathbf{q}(t))/\hbar} \quad (3)$$

where $\Gamma$ is usually set to be a diagonal width matrix with elements equal to the harmonic vibrational frequencies for bound state calculations. The pre-exponential factor is equal to

$$C_t(\mathbf{p}_0,\mathbf{q}_0) = \sqrt{\det\left[\frac{1}{2}\left(\mathbf{M_{qq}} + \Gamma^{-1}\mathbf{M_{pp}}\Gamma + \frac{i}{\hbar}\Gamma^{-1}\mathbf{M_{pq}} - i\hbar\Gamma\mathbf{M_{qp}}\right)\right]} \quad (4)$$

where $\mathbf{M_{ij}} = \partial \mathbf{i}_t/\partial \mathbf{j}_0$, $\mathbf{i}, \mathbf{j} = \mathbf{p}, \mathbf{q}$ is itself a matrix which represents a generic element of the monodromy (stability) matrix.[58, 69] The accuracy of the symplectic time evolution of the classical trajectories is monitored by checking at each timestep the deviation of the determinant of the positive-definite matrix $\mathbf{M}^T\mathbf{M}$ from unity[58].

Unfortunately, Monte Carlo integration of eq. (2) requires many thousands of trajectories to converge and it does not represent a feasible approach for complex molecular systems. A time-averaging filter was then introduced by Miller and Kaledin[70] with the goal to reduce this computational overhead. An additional approximation of eq. (4), called the separable approximation, consisting in substitution of the prefactor with its phase $C_t(\mathbf{p}(0), \mathbf{q}(0)) \approx e^{i\phi_t}$ where $\phi_t = \text{phase}[C_t(\mathbf{p}(0), \mathbf{q}(0))]$, allowed them to obtain the following expression for spectra calculations

$$I(E) = \left(\frac{1}{2\pi\hbar}\right)^F \iint d\mathbf{p}(0)d\mathbf{q}(0) \frac{1}{2\pi\hbar T} \left| \int_0^T e^{\frac{i}{\hbar}[S_t(\mathbf{p}(0),\mathbf{q}(0))+Et+\phi_t]} \langle \chi|\mathbf{p}(t),\mathbf{q}(t)\rangle dt \right|^2 \quad (5)$$

eq. (5) is the separable semiclassical time-averaged initial value representation (TA-SC-IVR) approximation for power spectra calculations. It does exploit an additional time integration to reduce the number of classical trajectories needed for the phase space integration. Equation (5) is also quite computationally convenient, since the phase integration is performed for a positive-definite integrand. It is found to be quite accurate over a wide range of molecular applications.[71-73] In particular, more recently, [74-76] a further approximation has been introduced that allows a reduction to just a handful of trajectories (or even a single one) per degree of freedom to converge eq. (5). This approximation is called multiple coherent SC-IVR (MC SC-IVR), since it is based on choosing the reference state $|\chi\rangle$ in eq. (5) as a tailored combination of coherent states, conveniently placed at the classical phase space points $(\mathbf{p}_{eq}^i, \mathbf{q}_{eq}^i)$, i.e. $|\chi\rangle = \sum_{i=1}^{N_{states}} |\mathbf{p}_{eq}^i, \mathbf{q}_{eq}^i\rangle$. We choose $\mathbf{q}_{eq}^i$ to be the equilibrium position vector, while $\mathbf{p}_{eq}^i$ is set in a harmonic fashion as $(p_{j,eq}^i)^2/2 = \hbar\omega_j(n_j + 1/2)$, where $j$ is a generic mass-scaled normal mode and $\omega_j$ is the associated frequency. Results of this approximation are quite accurate and MC SCIVR is a fairly good and trusty choice when the full phase space integration of eq. (5) is not doable, as in a complex system such as glycine.[71, 72, 77-83]

**III. Implementing the Semiclassical Initial Value Representation Method into VENUS**

The SC-IVR method was integrated into the general chemistry computer program VENUS[5, 6] to calculate the semiclassical vibrational power spectrum directly, while keeping all VENUS original features and introducing the least modifications possible into the code. The integration of SC-IVR into VENUS is shown in the flowchart of Figure 2. VENUS has been widely used in atomistic scale classical trajectory simulations for chemical reactions, and customized potential energy surfaces can be built from a wide set of potential energy functions to evaluate the potential energy and the gradient analytically in Cartesian space. Additional options for configuring SC-IVR calculations have been also integrated into the standard VENUS input file, including parameters for reference state, power spectrum, and deviation threshold of the monodromy matrix determinant from unity, as described in Figure 2. Besides analytic potential energy calculations, VENUS routines are retained for time evolution and trajectory information output.

The SC-IVR calculations are performed with the following scheme: 1) initial state sampling; 2) classical trajectory integration and semiclassical quantities calculation; and 3) power spectrum calculation at the end of each trajectory. The Hamiltonian trajectories are integrated with a 4$^{th}$ order symplectic integrator,[84-86] where the action and the monodromy matrix are updated at each sub-step required by the symplectic algorithm on the basis of momenta and potential energies, and numerically evaluated Hessian matrices, respectively. At each step, the semiclassical quantities, such as the time dependent prefactor and the coherent state overlap, are evaluated instantaneously, and a check of the deviation from unity of the determinant of the positive-definite matrix $\mathbf{M^TM}$ follows. The semiclassical trajectories are discarded once the deviation is larger than the defined threshold and they do not contribute to the spectrum calculation.

Basically, these procedures can be adapted to any classical or quasi-classical molecular dynamics code or package. The classical or quasi-classical trajectories are integrated with ordinary Newtonian or Hamiltonian methods, and two ways may be employed to calculate the semiclassical quantities. Either calculation is performed instantaneously as introduced above, or trajectories providing stepwise information are stored, and semiclassical quantities are calculated afterwards. The first way is straightforward, even though it requires modification of the original VENUS code and demands extra computer memory. The latter does not require code modification and allows reuse of data to perform SC-IVR calculations multiple times, when the trajectory calculation is computationally very expensive. However, issues might arise because of data storage and I/O requirements for large scale trajectory calculations.

**IV. Methodology**

The full dimensional and global potential energy surface (PES) employed for the semiclassical power spectra calculations of the $Cl^-$---$CH_3Cl$ pre-reaction complex was obtained by refining an ab initio analytical surface with experimental data.[31] The PES was implemented into the VENUS suite of codes[5, 6] and it was tested over the years with classical trajectory simulations.[25-31] In particular, rate constant calculations proved the PES to be quite reliable.[31] The surface is written in terms of internal coordinates, and it describes the reactant $Cl^-$ approaching the $CH_3Cl$ moiety. As $Cl^-$ and $CH_3Cl$ approach, the potential energy decreases because of the stretching and angular deformations induced by the attractive ion-dipole and ion-induced dipole forces. However, a pre-reaction complex is formed when the nucleophile and the carbon are bonded. The classical dissociation energy of the pre-reaction complex is 10.32 kcal/mol. The leaving group needs to surmount a barrier of 13.92 kcal/mol to leave the complex and complete the reactive process. A bi-dimensional cut of the PES employed is reported in Figure 3, showing the pre-reaction well. The vibrational energy levels of this multidimensional well are calculated and analyzed for spectroscopic features hinting at couplings between the vibrational modes.

To calculate the spectra we determined the full dimensional phase space integral of eq. (5) by sampling the initial classical trajectory positions around one of the two minima in Figure 3, and the momenta sampled around the harmonic zero point energy value with a Box-Muller method[87] according to a Husimi distribution[70]. The trajectory time-dependent positions are reported in color in Figure 3. None of the sampled trajectories escaped the potential well resulting in cluster dissociation or ending up in the other potential well (isomerization). Thus, the trajectories are bound and may be employed for vibrational spectra calculations of the pre-reaction well (provided the accuracy of the determinant of the monodromy matrix along the trajectory is sufficiently accurate for semiclassical calculation).

The most intuitive and direct way to evaluate the amount of coupling between modes is to look at the Hessian matrix off-diagonal elements at the classical PES global minimum. In normal mode coordinates, as in our case, these couplings are zero. In this case, then, one can either look at the higher potential derivatives, as in vibrational perturbation methods,[88] or calculate the Hessian at other representative geometries and take an average of the type

$$\bar{H}_{ij} = \sum_{k=1}^{N} |H_{ij}(t_k)|/N \qquad (6)$$

where $N$ is the number of sampled PES locations and $H_{ij}(t_k)$ is the $ij$ matrix element at the time-step $t_k$. We believe that a very representative collection of Hessian matrix elements is the one calculated during a classical trajectory starting at the global minimum geometry with a kinetic energy equal to the harmonic vibrational zero-point energy. The harmonic initial conditions bias the sampled PES somewhat above the actual quantum mechanical anharmonic zero-point energy and include anharmonicity effects typical of fundamental and lower overtone vibrational dynamics. Two modes are considered to be coupled if $\bar{H}_{ij} > \epsilon$, where $\epsilon$ is a fixed and arbitrary threshold value. If $\bar{H}_{ij} < \epsilon$ they are still considered coupled if there is a third mode $k$ such that both $\bar{H}_{ik} > \epsilon$ and $\bar{H}_{kj} > \epsilon$. In this case, all three modes, $i, j$ and $k$, are considered as coupled to each other and they are enrolled in the same "subspace", where a subspace with a collection of strongly interacting degrees of freedom is intended. Clearly, when $\epsilon = 0$ all modes are coupled, while for a big enough value of $\epsilon$ all modes are decoupled.[81, 83] In other words, to provide a complete coupling evaluation, different choices of $\epsilon$ must be tested.

A more accurate classical approach is to investigate energy flowing from one degree of freedom to another. This can be conveniently done by employing Liouville's theorem, which states that the infinitesimal phase space volume $d\mathbf{p}(t)d\mathbf{q}(t)$ is constant along a classical evolution or equivalently, that the Jacobian

$$\mathbf{J}(t) = \begin{pmatrix} \partial \mathbf{q}_t/\partial \mathbf{q}_0 & \partial \mathbf{q}_t/\partial \mathbf{p}_0 \\ \partial \mathbf{p}_t/\partial \mathbf{q}_0 & \partial \mathbf{p}_t/\partial \mathbf{p}_0 \end{pmatrix} \qquad (7)$$

is such that $\det(\mathbf{J}(t)) = 1$ at any time for the Hamiltonian (i.e. energy preserving) evolution. As an illustrative example, we consider two uncoupled degrees of freedom. Then, $\det(\mathbf{J}(t)) = \det(\tilde{\mathbf{J}}_1(t))\det(\tilde{\mathbf{J}}_2(t))$, where $\tilde{\mathbf{J}}_i(t)$ is the Jacobian of the $i$-th degree of freedom. In other words, if the group of modes are not coupled to each other, then the determinant of the full dimensional Jacobian is exactly equal to the product of the determinants of the sub-dimensional Jacobian matrices, because the full dimensional Jacobian can be written as a block diagonal matrix.[83] This observation can be very useful when trying to decipher the amount of coupling between modes. Upon setting the number *M* of degrees of freedom to be coupled out of the total *F* modes, one searches among all possible combinations of *M* modes from the larger set of *F* modes to find the *M*-dimensional subspace for the trajectory that has an average determinant of the Jacobian $\tilde{\mathbf{J}}_M(t)$ closest to unity.[83] We denote this approach as "Jacobi's criterion" from now on, and we will look for different sizes of mode coupling combination groups. The main drawback of Jacobi's criterion, with respect to the Hessian one, is that it has substantially a higher computational cost.

A quantum mechanical approach able to unveil the coupling between different vibrational modes can be obtained by inspecting the semiclassical power spectra calculated for different reference states $|\chi\rangle$. The reference state is usually written as the direct product of one dimensional coherent states

$$|\chi\rangle = |\mathbf{p}_{eq}, \mathbf{q}_{eq}\rangle = |p_{1,eq}, q_{1,eq}\rangle \ldots |p_{F,eq}, q_{F,eq}\rangle \quad (8)$$

and all types of vibrations are included. However, to enhance only spectral signals that correspond to eigenvalues with odd vibrational quantum number for a given *j*-mode and, at the same time, even quantum numbers for all the other modes, we can write the reference state in the following manner

$$|\chi\rangle = |\mathbf{p}_{eq}, \mathbf{q}_{eq}\rangle = (|p_{1,eq}, q_{1,eq}\rangle + |-p_{1,eq}, q_{1,eq}\rangle) \ldots (|p_{j,eq}, q_{j,eq}\rangle - |-p_{j,eq}, q_{j,eq}\rangle)$$
$$\times \ldots (|p_{F,eq}, q_{F,eq}\rangle + |-p_{F,eq}, q_{F,eq}\rangle)$$
$$= \prod_{j=1}^{M}(|p_{j,eq}, q_{j,eq}\rangle + \xi_j|-p_{j,eq}, q_{j,eq}\rangle) \quad (9)$$

as demonstrated by one of us for a set of harmonic oscillators.[76] $\xi_j$ is a parameter equal to $\pm 1$ according to the desired combination. In the example above, the peak is filtered to be the fundamental excitation of the $j$-th mode and the even excitations (or no excitations) of all the other modes. Thus, the allowed peaks will be given odd excitations of mode $j$ and even or no excitations of the other modes. Given the Husimi trajectory sampling, the fundamental excitations (1←0) are by far the most intense between all possible odd excitations. The first overtone (2←0) is the most intense for the even excitations. Thus, the possible peak combinations are given by $v_j + 2v_i$ for any $i \neq j$. However, only some of these mode combinations are intense enough to be detected. These are the ones for which couplings between modes $j$ and $i$ are stronger. Advantages of this approach are manifold. First, the results are not representative of a single trajectory simulation, as in the case of the averaged Hessian and the Jacobi's criterion, but of thousands of trajectories. Clearly, this amount of trajectories is more representative of the entire PES and of couplings between the modes. Secondly, it is a quantum mechanical approach and it includes the role played by quantum mechanical effects in energy exchange between modes. This is important for having novel insights into the well-studied nucleophilic reactions.

**V. Results and Discussion**

The goal of this section is to show how the methodology described in Sec. II may provide a deep physical quantum mechanical picture, in particular for the reaction mechanism of nucleophilic reactions. We adopt a spectroscopic approach and calculate the vibrational frequencies at different levels of approximation. We first report the classical power spectrum for the fundamental frequency estimates. Then we calculate several spectra using the semiclassical approximation of eq. (5) and obtain a quantum mechanical estimate of fundamental and overtone frequencies. Then, we compare the semiclassical results with another approximate quantum mechanical method, second-order Vibrational Perturbation Theory (VPT2).[88, 89] Finally, we use all this information and that obtained from the averaged Hessian and the Jacobi's criterion to understand vibrational couplings and their importance for the reaction mechanism.

## A. Classical power spectrum

In classical mechanics it is possible to have an estimate of mode frequencies by Fourier transforming the motion correlation function. An example is given by the Fourier transform of the velocity autocorrelation function

$$C_{vv}(t) = \langle v(t)v(0) \rangle \tag{10}$$

where the average $\langle \ldots \rangle$ is taken for a given ensemble of trajectories. In our case, for better comparison with the semiclassical results reported below, we employ the same ensemble of classical trajectories that we use for the phase space Monte Carlo calculation of eq. (5). This classical simulation has the advantage with respect to a single point Hessian diagonalization, i.e. to the harmonic estimate, to include classical anharmonic effects and mode resonances. For example, Fermi resonances can be detected also by classical simulations and they occur when a combination of mode frequencies approximately equals another mode frequency of the same symmetry. For example, when a mode j is such that its frequency $v_j \approx 2v_i$, the latter being the first overtone of another mode $i$, a Fermi resonance occurs and the $v_j$ fundamental signal is split into two peaks.[90] However, classical simulations do not include any zero point energy (ZPE) or tunneling effects. Also, quantum anharmonic and quantum resonance effects are not reproduced.

Figure 4 reports the Fourier transform of eq. (10) for the pre-reaction complex $Cl^-$---$CH_3Cl$. At low frequency values, one can detect the fingerprints of bending and stretching involving the weakly bound $Cl^-$, while at higher frequencies deformation of the molecular $CH_3Cl$ species is observed. A detailed assignment is reported further below in Table I and a discussion will follow.

**B. Semiclassical power spectra**

We now add quantum mechanical effects to our classical trajectories by using eq. (5) to calculate the power spectrum of the $Cl^-$---$CH_3Cl$ complex. As explained above, the choice of the reference state $|\chi\rangle$ allows for peak selection. When using a combination of the type given by eq. (9), in which $\xi_j = 1$ for each $j$, the ZPE peak and all even excitations are enhanced. This spectrum can be observed at the bottom of panel (a) of Figure 5 in black color. In the same figure spectra with $\xi_j = -1$ for each value of $j$ going from mode 1 to mode 12, as labeled in each plot, are reported. We show spectra obtained using both 200,000 (dashed line) and 500,000 (continuous line) trajectories. The signals are very similar so we deem that the simulations have reached Monte Carlo convergence. Every classical trajectory is propagated for 20,000 atomic time units (483.78 fs), employing a time step equal to 10 a.u.. This time length is close to typical values employed in previous semiclassical power spectra calculations.[72, 79] The consistency of our symplectic simulation and the accuracy of the pre-exponential factor in eq. (4) are guaranteed by checking that the determinant of the monodromy matrix is within a value of $10^{-3}$ from unity.

The main information from Figure (5) is the frequency of each fundamental. These values are reported in Table I. The first column of this table enumerates the modes, while in the second column the corresponding irreducible representation of the $C_{3v}$ point group of symmetry is listed. Eight modes belong to the E irreducible representation, while the remaining four to the total-symmetric $A_1$ irreducible representation. The description of each normal mode is reported in the third column. Column four gives the harmonic frequencies, which are commonly employed to evaluate the accuracy of a PES. However, the following columns show how harmonic values can be significantly corrected once the anharmonic effects are accounted for. It is interesting to note that the anharmonicity can shift the harmonic estimates both to lower or higher values. For example, the high CH stretch frequency modes are less stiff after the inclusion of anharmonic effects. Instead, the $CH_3$ deformation modes are stiffer. The classical simulation does account for anharmonicity, since the classical trajectories explore all the PES beyond the harmonic surroundings of the global minimum. Also, the classical simulation is able to catch normal mode resonances, which show up as beatings in the correlation functions. For example, in our case, the classical dynamics calculation shows how the two highest frequency CH stretch modes, which are degenerate at the global minimum according to the E irreducible representation, are split into two peaks due to the Fermi resonance occurring with the $CH_3$ deformation modes 7 and 8 which also belong to the E irreducible representation. In fact, the double of the fundamental frequency of modes 7 and 8 is not far from the CH stretch fundamental. This coupling is not observed at a harmonic level of theory. Thus, the vibrational picture is quite complicated and the modes are strongly coupled. This conclusion is supported by the observation in Figure 5 that included in the spectra are several peaks of strong intensity in addition to the fundamental mode. These peaks denote the presence of several overtones and peak splittings. A comparison with the classical picture of Figure 4 shows that the quantum mechanical vibrational motion of this pre-reaction complex is much more complicated than its classical counterpart. In fact, the numerous quantum overtones do not belong to a single mode but to combinations of different modes. This proves that the tentative standard classical picture of independent normal modes is far from being realistic.

From Table I we observe that for the semiclassical results all degeneracies are removed, differently from the classical simulation. We think that this is not necessarily due to quantum effects since the energy differences are very small and within semiclassical accuracy. An approximate comparison may be made with the experimental frequencies for the isolated $CH_3Cl$ molecule,[91] which are reported in the last column of Table I. Even if a direct comparison is not possible, we can appreciate similarities between the semiclassical and experimental values, and for both there is a Fermi resonance between the CH stretch modes and the $CH_3$ rocking modes. Another feature, that the comparison with the experiment points out, concerns the frequency of the C-$Cl_b$ stretch (mode 4). For the isolated molecule the anharmonic experimental frequency[91] of this mode is 733 cm$^{-1}$, and the harmonic frequency[31] calculated at HF/6-31G* theory level is 783 cm$^{-1}$. However, for the pre-reaction complex, the harmonic frequency is 662 cm$^{-1}$ and it is red shifted by an amount of about 120 cm$^{-1}$.[31] This peculiarity of the mode 4 frequency can be related to weakening of the C-$Cl_b$ bond when the C-$Cl_a$ one is created and the complex formed.

Finally, we compare the semiclassical results of Table I with other quantum vibrational methods. For this purpose, we performed VPT2[88, 89, 92] calculations using the MOLPRO suite of codes.[93, 94] We chose two levels of ab initio theory to compare with the PES employed in the dynamic calculations. The first VPT2 calculation was performed using HF/6-31G*, which is the level of theory for the ab initio points of the PES.[31] The results are reported in Table II in the fourth and fifth columns, respectively for the harmonic and VPT2 approximations. The second calculation was at the level of MP2/6-311++G** and the results are reported in the last two columns of the same Table. The most striking feature is that this vibrational quantum mechanical method is not able to quantitatively reproduce the Fermi resonance splittings observed both in the dynamics calculations described above and in the experiments.[91] This can be in part explained considering that the present VPT2 version is based on a perturbation calculation done at the global minimum, where the $C_{3v}$ symmetry enforces the double degeneracy. Newer VPT2 implementations, which are in the process of testing, are better suited for the Fermi resonance calculations.[95] VPT2 is quite accurate in accounting for anharmonic effects.[96-101] If one looks at the anharmonic shift in the VPT2 results at the level of MP2/6-311++G** calculation and compares it with the corresponding shifts of the classical $C_{vv}(t)$ and semiclassical simulations, there are many similarities. Specifically, for the two degenerate $CH_3$ deformation modes the quantum anharmonic corrections, i.e. both semiclassical and VPT2 ones, have opposite signs with respect to the classical anharmonicity. Conversely corrections are quantitatively quite similar in the case of the CH stretch modes. More quantitative comparisons are not possible due to the different level of theory employed for the PES and the VPT2 calculations.

## C. Understanding couplings between the vibrational modes

We now focus on the investigation of couplings between the vibrational modes of the pre-reaction complex. The goal is to reach a deep and quantum mechanical physical insight into the nucleophilic reaction mechanism. In previous publications,[25-29] classical trajectories with different initial conditions have been the main tool for investigation of vibrational couplings and the reaction mechanism. In the following we employ three alternative approaches, outlined in the methodology Section IV.

The first approach is based on the Hessian matrix of eq. (6) averaged over a classical trajectory started from the global minimum, i.e. the most probable trajectory in the semiclassical calculation described above. We track the subspace composition for different coarse-grained parameters $\epsilon$. Figure 6 shows the trend of the maximum averaged Hessian matrix subspace dimension versus different values of $\epsilon$. If compared with other molecules,[81, 83] the staircase profile of Figure 6 and the gradual decrease in dimensionality suggests that almost all modes are coupled to each other. For $\epsilon = 0$ all 12 vibrational modes are in the subspace and as $\epsilon$ is increased, modes are removed. The first mode removed is the $Cl_a^-$---$CH_3Cl$ stretching, and the next two are the degenerate $Cl_a^-$---$CH_3Cl$ bendings, reducing the subspace dimension to 9. At the smallest $\epsilon$ value, i.e. in the upper part of the stair, we observe that the less coupled mode is mode 3, the C-$Cl_a$ stretching, as expected since the bond is weak, i.e. $Cl^-$---$CH_3Cl$. By increasing the parameter $\epsilon$, i.e. by decreasing the graining, the other two modes related to the $Cl_a$ motion, i.e. the bending ones labeled by 1 and 2 in Table I, are shown to be less coupled. From these considerations we come to a first conclusion which is that the bending motions involving $Cl_a$ are more coupled to the $CH_3Cl_b$ molecule than the C-$Cl_a$ stretch. This is directly related to the observation that the $Cl^-$---$CH_3Cl$ unimolecular dynamics with excited intermolecular modes are not affected in the short time by the $CH_3Cl$ vibrational normal modes.[24] The intermolecular vibrational frequencies are 70, 71 and 111 cm$^{-1}$ for the two C-$Cl_a$ bendings and for the C-$Cl_a$ stretching, respectively, and they are much smaller than the vibrational frequencies of $CH_3Cl_b$. The uncoupling between the intermolecular vibrational modes and the intramolecular ones results in inefficient intramolecular vibrational energy redistribution, i.e. there is a bottleneck that prevents excess energy in the $CH_3Cl_b$ modes to flow into the three C-$Cl_a$ modes. This non-RRKM behavior is also seen in the scattering of $Cl^-$ with vibrationally excited $CH_3Cl$ and in the long lifetime of the $Cl^-$---$CH_3Cl$ pre-reaction complex with $CH_3Cl$ vibrational excitation.[23,26] Even if the semiclassical $Cl^-$---$CH_3Cl$ trajectories have on average an energy of 24.34 kcal/mol, which is well above the transition state barrier height of 13.92 kcal/mol from the potential minimum, the inefficient intramolecular vibrational energy redistribution leads to a long lifetime suitable for SC-IVR spectroscopy calculations. If we keep on trying to break down the full-dimensional vibrational space into even smaller subspace, we find that the next mode to be separated in this artificial uncoupling procedure is the C-Cl$_b$ stretch, leaving the methyl group vibrations into one subspace. This suggests that Cl$_b$ is somehow more coupled to the methyl group than Cl$_a$, which is expected. Finally, all one-dimensional subspace are found for a big enough $\epsilon$ value, i.e. $\epsilon > 10^{-6}$.

A more sophisticated method to investigate vibrational couplings is the Jacobi's criterion, described in Section IV. Figure 7 reports the deviation from unity of the determinant of the reduced dimensionality $M$ Jacobian $\det\left(\tilde{J}_M(t)\right)$ for the same classical trajectory employed in the case of the averaged Hessian procedure reported above. Clearly, there is no deviation from unity of the determinant in Figure 7 for $M = 12$, since in this case the Jacobian is full dimensional and Liouville's theorem is fully satisfied. Lowest deviations from unity can be found for subspace of dimensionality $M = 11$ and $M = 9$, or equivalently $M = 3$ since these are the three degrees of freedom left out by the 9-dimensional subspace. In the case of $M = 11$, the subspace is composed of all degrees of freedom with exception of the C-Cl$_a$ stretch, in agreement with the averaged Hessian results. When the breakdown of the vibrational space into two subspaces of dimension 9 and 3 is enforced, we find that the 9-dimensional space is composed of all the modes related to the CH$_3$ vibrational motion with the addition of one C-Cl$_a$ bending. This is a key result, since it shows once again that in terms of energy flowing from one mode to another, it is one C-Cl$_a$ bending which is responsible for the transfer of the incoming collisional energy to the methyl group leading to nucleophilic substitution. The Jacobi's criterion shares with the Hessian method the importance of the C-Cl$_a$ bending mode for the reaction mechanism and differs from it in suggesting the C-Cl$_b$ stretch to be less coupled to the methyl group than the C-Cl$_a$ bending ones. If $M$ is set equal to 8, then we find the eight methyl group vibrational modes as the best eight-dimensional Jacobi subspace.

Both previous procedures are classical and based on a single trajectory. We aim at including quantum mechanical effects and a statistical average over thousands of trajectories for our vibrational coupling study. For this reason, we look at the semiclassical spectra plotted in Figure 5. The overtone peaks of each partial spectra clearly suggest some kind of coupling between the modes. However, a definitive assignment of each peak is quite cumbersome, due to the complexity of the system. Since we are interested on how the low frequency modes, which are related to the $Cl_a$ motion, are coupled to the other modes, we examine in Figure 5 couplings between these modes and the $CH_3Cl_b$ modes. It is clear from Figure 5 that the spectrum providing the fundamental of mode 6, one of the $CH_3$ rocking modes, shows many low frequency overtones. This suggests that a key role in activation of the nucleophilic reaction is played by the $CH_3$ rocking vibrational mode. It is interesting that the other $CH_3$ rocking mode, mode 5, is not as strongly coupled to other modes as mode 6. Figure 5 also shows that the high frequency modes, i.e. $CH_3$ deformations and CH stretches, are much less coupled to the low frequency modes. This is also in agreement with the previous findings that the mode specific excitation of the $CH_3$ deformation does not promote the $S_N2$ reaction[25-31] and excitation of CH symmetric stretching modes is ineffective to $S_N2$ reaction as shown in a recent experimental study.[102] We stress once more that this conclusion has been reached with a quantum mechanical approximate method and employing 500,000 classical trajectories.

**VI. Conclusions and Outlook**

In this work, we have presented implementation of SC-IVR MD for spectroscopic calculations on the general chemical dynamics computer program VENUS. We have shown in Section III how this may be done by retaining as much as possible the original structure of the code. This user-friendly implementation of the SC-IVR method may be performed with other existing codes and it opens up the possibility for users who are not experts of semiclassical methods to employ SC-IVR dynamics.

The new SC-VENUS suite of codes was applied to the calculation of the vibrational energy levels of the pre-reaction $Cl^-$---$CH_3Cl$ complex. SC-VENUS provides a quantum mechanical picture of intermode coupling, while two other methods, i.e. the averaged Hessian subspace decomposition and Jacobi's criterion, are purely classical and have also been employed for the same purpose. A common picture emerges from the three different approaches. The vibrational degrees of freedom mostly involved in the nucleophilic substitution are those related to the incoming projectile $Cl_a^-$ and the $CH_3$ rocking. More specifically, the $Cl_a^-$ reactive motion is given by the degenerate free rotation around the $CH_3Cl_b$ molecule when it is far apart, and by the degenerate C-$Cl_a$ bending modes in the pre-reaction $Cl_a^-$---$CH_3Cl_b$ complex. It is steering of the kinetic energy of one of these modes that activates the $CH_3$ rocking and promotes nucleophilic substitution. This conclusion was in part suggested in a previous publication,[31] where it was observed that the $CH_3$ rocking mode changes a lot along the reaction path. A significant role in this steering orientation is given by the chloromethane dipole moment which orients the reactants, especially at low temperatures.[29] However, the quantum mechanical picture of the semiclassical calculations is less neat and it suggests that there are probably several paths that can contribute to promote the $CH_3$ rocking − C-$Cl_a$ bending energy exchange, which we conclude to be the core of the reaction mechanism.

Additionally, the SC-IVR spectroscopic calculation provides anharmonic vibrational zero-point energy with errors within a few wavenumbers, which is essentially very accurate for the anharmonic zero-point energy correction in thermochemistry. One may perform such an SC-IVR calculation with much fewer trajectories or with just the most probable trajectory to calculate the quantum anharmonic zero-point energy of a molecule. This is more practical and reliable when comparing thermochemical data with experimental measurements than using the harmonic zero-point energy.

**Acknowledgment**

Michele Ceotto, Giovanni Di Liberto and Riccardo Conte acknowledge financial support from the European Research Council (ERC) under the European Union's Horizon 2020 research and innovation programme (Grant Agreement No. [647107] – SEMICOMPLEX – ERC-2014-CoG). M.C. acknowledges also the CINECA and the Regione Lombardia award under the LISA initiative (grant GREENTI) for the availability of high performance computing resources. Xinyou Ma and William L. Hase acknowledge financial support from the National Science Foundation under Grant No. CHE-1416428, the Robert A. Welch Foundation under Grant No. D-0005, and the Air Force Office of Scientific Research under AFOSR Award No. FA9550-16-1-0133. The development of VENUS-SCIVR and preliminary tests were performed on the Chemdynm computer cluster of the Hase research group, while the $Cl^-$---$CH_3Cl$ trajectory and spectra calculations were performed at the High Performance Computing Center (HPCC) at Texas Tech University (TTU).

**Table I.** Vibrational Fundamental Transitions for the Cl$^-$---CH$_3$Cl Pre-reaction Complex$^a$

| Mode | Irr. Repr. | Type | Harmonic | Classical | SC-IVR 200k | SC-IVR 500k | expt. CH$_3$Cl |
|---|---|---|---|---|---|---|---|
|  |  | ZPE | 8513 |  | 8435 | 8435 |  |
| 1 | E | Cl$_a$−C bend | 72 | 76 | 53 | 70 |  |
| 2 | E | Cl$_a$−C bend | 72 | 76 | 73 | 71 |  |
| 3 | A$_1$ | Cl$_a$−C stretch | 107 | 111 | 123 | 111 |  |
| 4 | A$_1$ | Cl$_b$−C stretch | 662 | 647 | 633 | 625 | 733 |
| 5 | E | CH$_3$ rock | 1131 | 1111 | 1100 | 1098 | 1017 |
| 6 | E | CH$_3$ rock | 1131 | 1111 | 1110 | 1104 | 1017 |
| 7 | E | CH$_3$ deformation | 1462 | 1484 | 1451 | 1453 | 1452 |
| 8 | E | CH$_3$ deformation | 1462 | 1484 | 1459 | 1455 | 1452 |
| 9 | A$_1$ | CH$_3$ deformation | 1555 | 1550 | 1527 | 1544 | 1355 |
| 10 | A$_1$ | CH stretch | 3036 | 2974 | 2854 | 2870 | 2879 |
| 11 | E | CH stretch | 3166 | 3053 | 3058 | 3064 | 2968 |
| 12 | E | CH stretch | 3166 | 3107 | 3033 | 3078 | 3039 |

*a.* Frequencies are in cm$^{-1}$ and for the PES used in the simulations. The first column enumerates the modes, the second reports the corresponding irreducible representation of the C$_{3v}$ symmetry group and the third the type of normal mode. In the following columns the frequencies respectively in harmonic, classical, and semiclassical (with 200,000 and 500,000 trajectories) approximation are reported. The last column are the experimental values for the isolated CH$_3$Cl molecule.[91]

**Table II.** Vibrational Fundamental Transitions for the Pre-Reaction Complex Cl$^-$---CH$_3$Cl Using Vibrational Second-Order Perturbation Theory (VPT2)$^a$

| Mode | Irr. Repr. | Type | HF/6-31G* Harmonic | HF/6-31G* VPT2 | MP2/6-311++G** Harmonic | MP2/6-311++G** VPT2 |
|---|---|---|---|---|---|---|
|  |  | ZPE | 8998.85 | 8887.06 | 8500.27 | 8401.71 |
| 1 | E | Cl$_a$–C bend | 71.44 | 73.22 | 56.76 | 82.87 |
| 2 | E | Cl$_a$–C bend | 71.44 | 71.00 | 56.76 | 83.25 |
| 3 | A$_1$ | Cl$_a$–C stretch | 100.83 | 101.48 | 103.94 | 104.97 |
| 4 | A$_1$ | Cl$_b$–C stretch | 653.66 | 634.33 | 707.30 | 692.00 |
| 5 | E | CH$_3$ rock | 1100.25 | 1081.71 | 1023.51 | 1008.04 |
| 6 | E | CH$_3$ rock | 1100.25 | 1082.09 | 1023.51 | 1007.94 |
| 7 | E | CH$_3$ deformation | 1605.52 | 1569.29 | 1462.24 | 1436.41 |
| 8 | E | CH$_3$ deformation | 1605.52 | 1569.31 | 1462.25 | 1436.73 |
| 9 | A$_1$ | CH$_3$ deformation | 1477.76 | 1452.09 | 1385.65 | 1348.07 |
| 10 | A$_1$ | CH stretch | 3320.29 | 3224.70 | 3159.49 | 3044.68 |
| 11 | E | CH stretch | 3445.37 | 3308.26 | 3279.56 | 3136.26 |
| 12 | E | CH stretch | 3445.37 | 3308.32 | 3279.57 | 3135.88 |

*a*. Frequencies are in cm$^{-1}$. The first column enumerates the modes, the second reports the corresponding irreducible representation of the C$_{3v}$ symmetry group, and the third the type of the normal mode. In the fourth and fifth column ab initio frequencies are reported at the HF/6-31G* level with the harmonic approximations and using the VPT2 method. The last two columns are the same but at the MP2/6-311++G** level.

**Figure Captions**

**Figure 1.** The pre-reaction complex Cl$^-$---CH$_3$Cl. Green for chlorine atoms, brown for carbon and gray for hydrogen atoms.

**Figure 2.** Algorithm flowchart illustrating the implementation of the SC-IVR subroutines into VENUS. The red boxes represent the SC-IVR subroutines, whereas the green boxes represent the modified VENUS subroutines.

**Figure 3.** Accepted trajectories for the semiclassical calculations are plotted onto the potential energy surface contour map. The coordinates are the two C-Cl bond distances as coordinates. The dynamics of all accepted trajectories, represented by the colored lines, is confined within a potential well and neither dissociation nor isomerization were observed.

**Figure 4.** Classical power spectrum for the pre-reaction complex Cl$^-$---CH$_3$Cl from the Fourier transform of the velocity autocorrelation function.

**Figure 5.** Fundamental excitations of the pre-reaction complex Cl$^-$---CH$_3$Cl, selected by tailoring the reference state $|\chi\rangle$ according to eq. (9). Continuous lines are for the semiclassical simulations with 500,000 trajectories, while dashed ones for the 200,000-trajectory simulations. Panel (a) reports the fundamentals belonging to the A$_1$ irreducible representation of the C$_{3v}$ symmetry group, while panel (b) shows excitations belonging to the E irreducible representation. The ZPE signal is in black at the bottom of panel (a). Labels are assigned according to the harmonic frequencies reported in Table I. There is no relation between same colors in the two panels.

**Figure 6.** Trend of the highest dimensional subspace of the averaged Hessian matrix for different values of the coarse-grained parameter $\epsilon$. The picture can be interpreted as the number of modes that remain coupled while increasing the threshold $\epsilon$.

**Figure 7.** Average values of $\left|1 - \det\left(\tilde{\mathbf{J}}_M(t)\right)\right|$ for subspaces of different dimensionality (M).

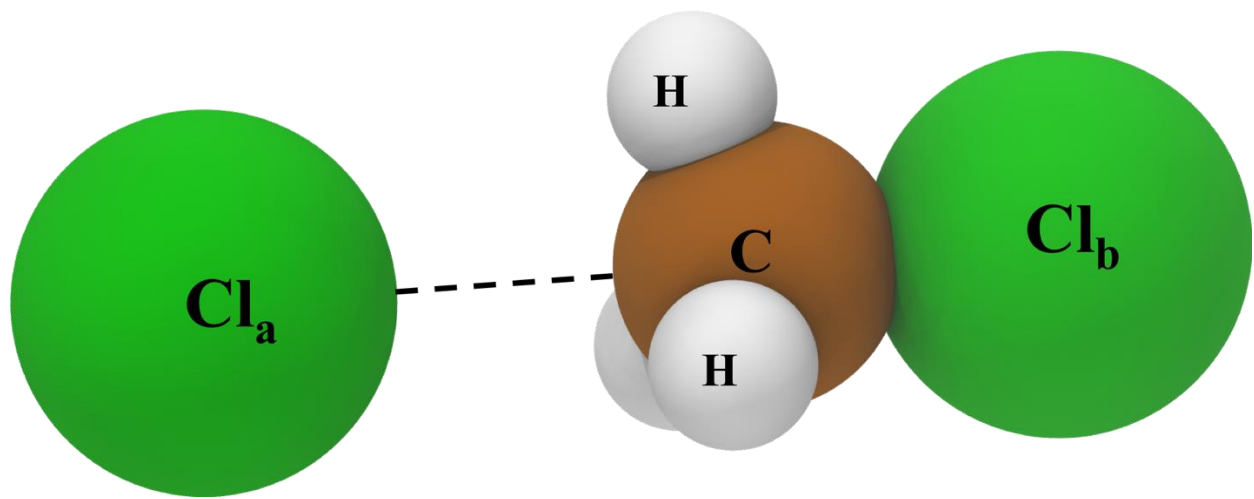

**Figure 1.**

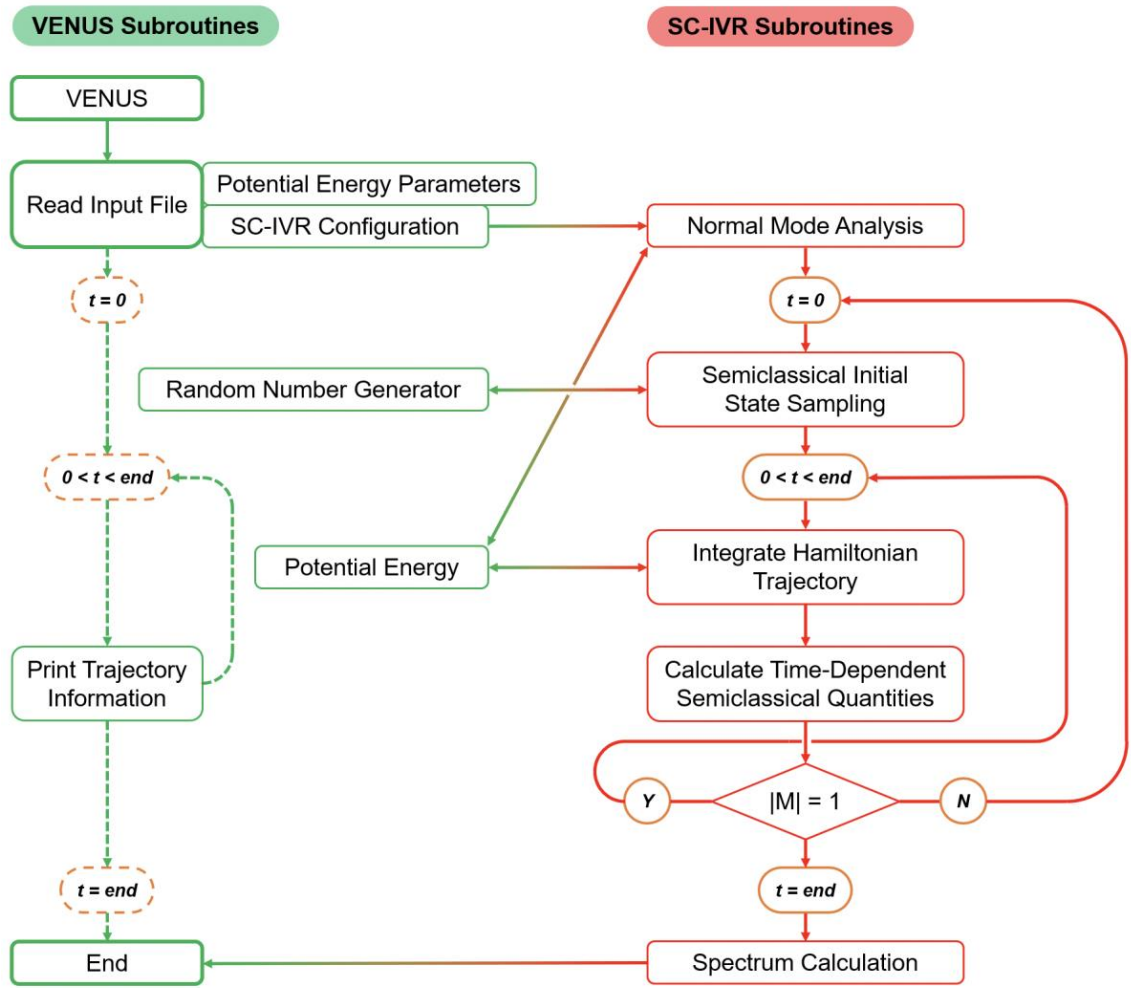

**Figure 2.**

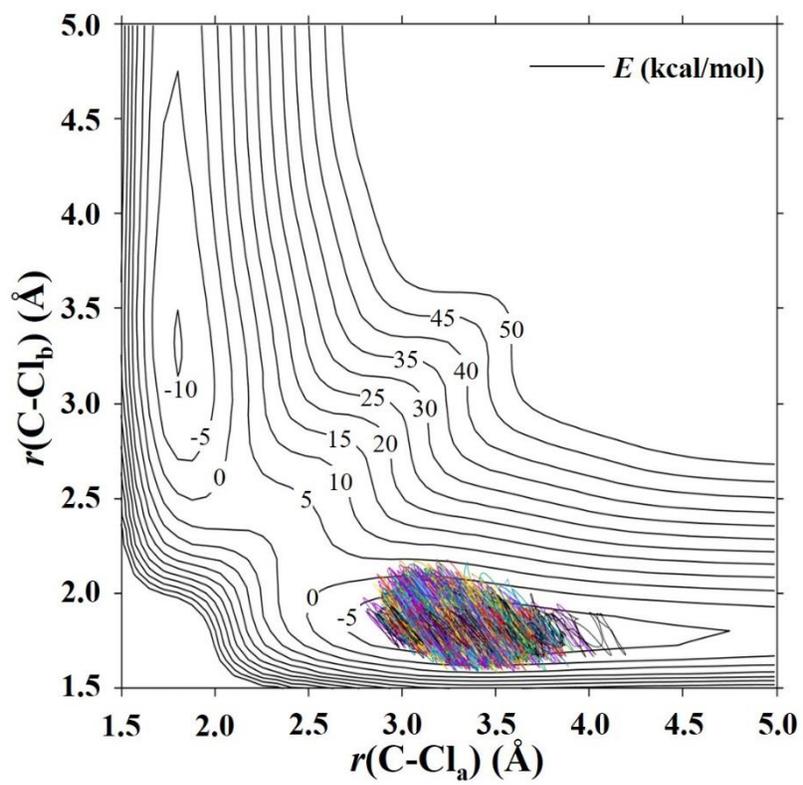

**Figure 3.**

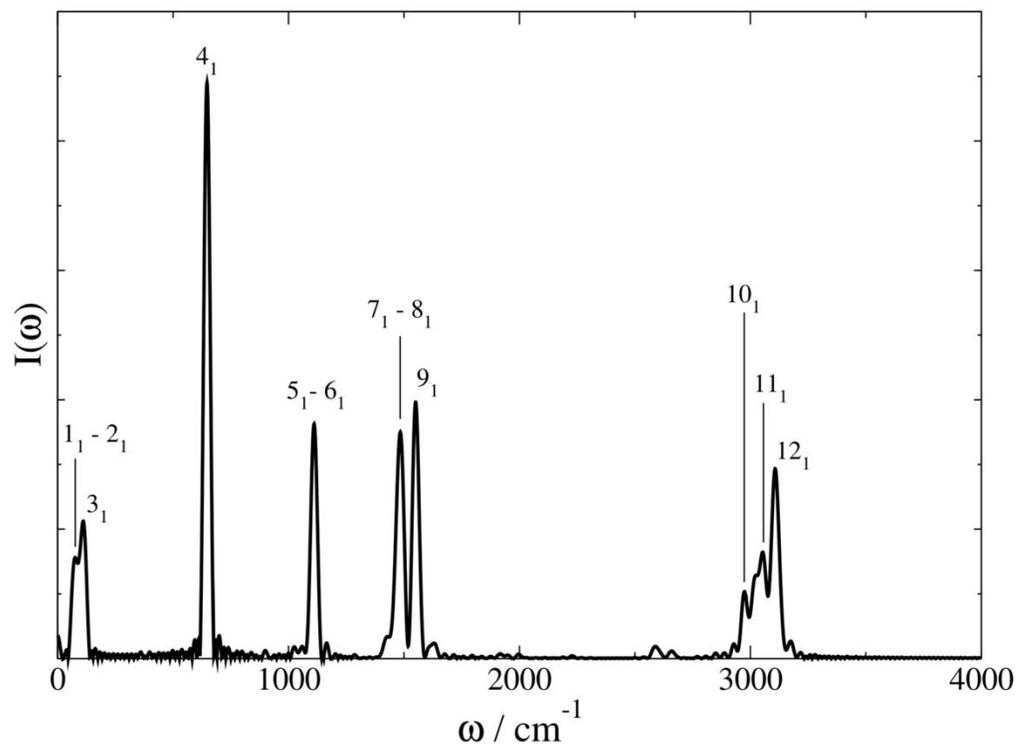

**Figure 4.**

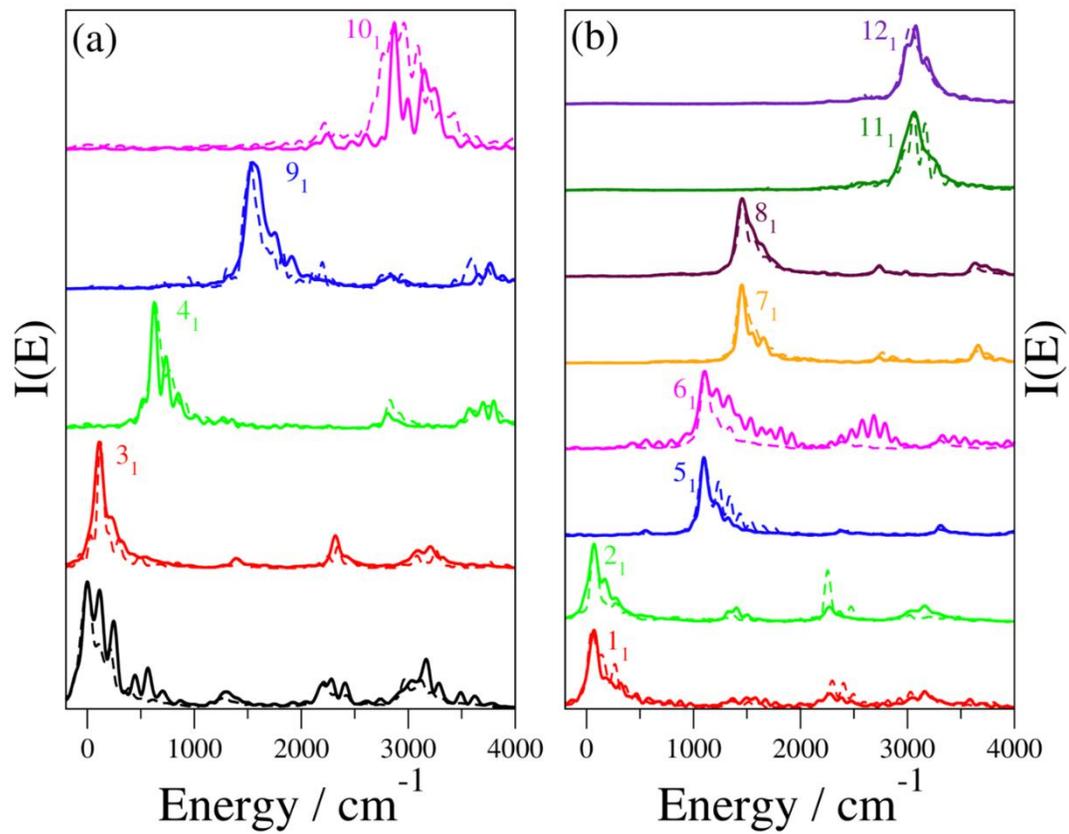

**Figure 5.**

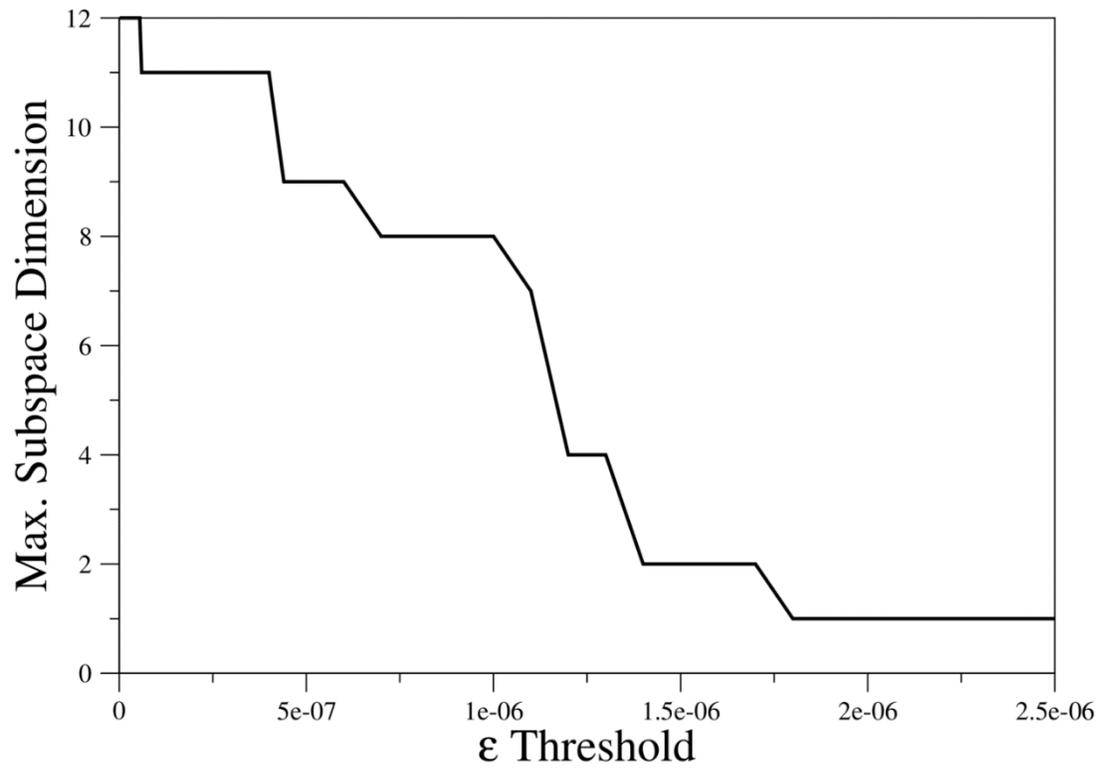

**Figure 6.**

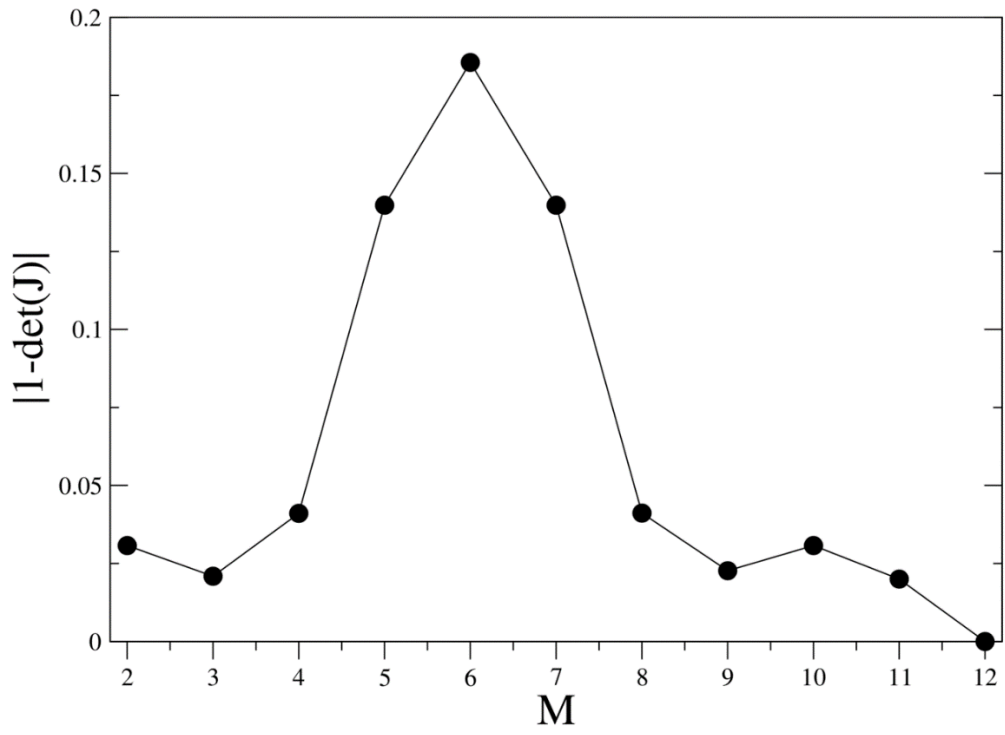

**Figure 7.**